\documentclass[DIV12]{scrartcl}

\usepackage{hyperref}
\usepackage[utf8]{inputenc}
\usepackage{tabularx}
\usepackage{setspace}
\usepackage{graphicx}
\usepackage{booktabs}
\usepackage{amsmath}
\hypersetup{
	colorlinks=true,
	linkcolor=black,
	citecolor=black,
	urlcolor=black
}

\date{}
\usepackage{authblk}
\usepackage{blindtext}

\usepackage{natbib}
\usepackage{fancyhdr}
\begin{document}

\title{Machine listening intelligence}

\author[1]{Carmine-Emanuele Cella}
\affil[1]{\small Ircam, Paris, France}

\maketitle

\thispagestyle{fancy} 

\begin{abstract}
This manifesto paper will introduce \emph{machine listening intelligence,
}an integrated research framework for
acoustic and musical signals modelling, based on signal processing,
deep learning and computational musicology.
\end{abstract}


\section{Introduction}

\subsection{Motivation}

The relation between \emph{signals }and \emph{symbols} is a central
problem for acoustic signal processing. Among different kind of signals, musical
signals are specific examples in which there is some information regarding
the underlying symbolic structure. While an impressive amount of research has been done in
this domain in the past thirty years, the symbolic
processing of acoustic and musical signals is still only partially
possible. 

The aim of this paper, grounded on our previous work \citep{Cella2009}, is
to propose a manifesto for \textbf{a generalised approach for the representation of
acoustic and musical signals called}\textbf{\emph{ machine listening
intelligence}}\textbf{ (MLI), }by integrating cognitive musicology
insights, hand-crafted signal processing and deep learning methods
in a general mathematical framework. 

Among existing approaches that share similarities with ours, there are 
the \emph{multiple viewpoint system} 
\citep{Conklin2013} and \emph{IDyOM} \citep{Pearce2012}. While comparing differences and
similiarities with these approaches could be interesting, we will not do this 
here and we mention them only for reference.

\subsection{Scientific assessment}

In the past twenty years, with the improvement of computers and the
advancements in machine learning techniques, a whole field called
\emph{music information retrieval (MIR) }developed massively. This
important domain of research brought impressive results and had been
able to tackle problems that appeared to be unsolvable, such as the
classification of very complex signals. Nonetheless, 
tasks that are relatively easy for humans
are still hard and there are not general solutions. Apparently, 
these kind of tasks are often ill-defined or lack proper
information to be correctly represented.

More recently, the uprise of deep learning techniques in computer
vision created a real revolution in machine learning given the advancements
they provided \citep{Krizhevsky2012}. Deep convolutional networks,
for example, provide state of the art classifications and regressions
results over many high-dimensional problems \citep{LeCun2015}. Their
general architecture is based on a cascade of linear filter weights
and non-linearities \citep{Mallat2016}; the weights are learned from
massive training phases and generally outperform hand-crafted features. The switch to
large scale problems that happened in the past few years in the computer
vision community, moreover, proved the fact that we need to address
more general problems for acoustic signals, where the domain of research
is not defined by a single sample but by a whole \emph{population}. 

However, these complex programmable machines bring us to a very difficult and partially
unknown mathematical world.  We believe that \emph{representation theory}
is good candidate for such mathematical model. Signal representation
methods can be thought as linear operators in vector space and representation
theory studies abstract algebraic structures by representing their
elements as linear transformations of vector spaces \citep{Fulton2004}.

Next sections will be as follows: section \ref{sec:Problems-and-outcomes}
will show some problems and applications that we would like to address
with our approach. From section \ref{sec:Properties-of-a} we will
go more into technical details describing what are representations
for signals and reviewing their properties both from cognitive and
mathematical standpoints. This will serve a background for
section \ref{sec:researchapproach}, that will present the general research
approach for MLI.

\section{Problems and outcomes\label{sec:Problems-and-outcomes}}

Among the large number of open problems in the field of signal processing,
we would like to present here some interesting examples that could
be treated in the context of machine listening intelligence. These
problems refer particularly to music and creative applications, but
we think that the developed methodology could be further used in different
domains. As such, they must be considered just as examples of possible
outcomes.

\subsection{Semantic signal processing}

A signal transformation is, in a general sense, any process that changes
or alter a signal in a significant way. Transformations are closely
related to representations: each action is, indeed, performed in a
specific representation level. For example, an elongation performed
in time domain gives inferior results (perceptually) to the same transformation
performed in frequency domain, where you have access to phases. In
the same way, a pitch shift operated in frequency domain gives inferior
results to the same operation performed using a spectral envelope
representation, where you have access to dominant regions in the signal.
In the two cases discussed above we passed from a low-level representation
(waveform) to a middle-level representation (spectral envelope). We
could, ideally, iterate this process by increasing the level of abstraction
in a representation thus giving access to specific properties of sound
that are perceptually relevant; by means of a powerful representation
it could therefore be possible to access a \emph{semantic level }for
transformations. We envision, therefore, the possibility in the future to have such
kind of semantic transformations by accessing representations given
by deep learning models.

\subsection{High-level acoustic features}

One of the most impressive example of creative application of deep
learning is, in our opinion, style transfer. In a paper published
in 2016 \citep{Gatys2016}, for example, the authors showed how it
is possible, using the latent space of a deep network, to transfer
high level features from one image to another. It is interesting to
remark that the transferred features are not simple \emph{effects}
but real traits of the style of the image.

Unfortunately, such an impressive process is not possible on acoustic
signals for the moment. In 2013 we achieved the construction of an
advanced hybridisation system for sounds \citep{Cella2013}. The basic
idea of our approach was to represent sounds in term of classes of
equivalences and probabilities, then mix the classes of one sound
with the probabilities of another one. While the perceptual results
were satisfying for us, the limitation of the representation method
used, didn't permit to access \emph{high-level acoustic features}.

We believe that deep musical style transfer can be considered as a generalisation
of hybridisation and we strongly believe that this kind of processing
could be achieved by MLI.

\section{Signal representations\label{sec:Properties-of-a}}

Defining a representation for music and musical signals involves the
establishment of essential properties that must be satisfied. Many
years of research have been devoted to such a complex task in the
field of cognitive musicology, a branch of cognitive sciences focused
on the modelling of musical knowledge by means of computational methods
\citep{Laske1992}. 

Among important properties for musical representations
found from the literature in the field, there are milestones such as
\emph{multiple abstraction levels}, \emph{multi-scale} and \emph{generativity}.
Interestingly enough, these properties are not only specific to music
but can be applicable to different kind of acoustic signals, as we
will see. 

\begin{figure}
\centering{}\includegraphics[scale=0.45]{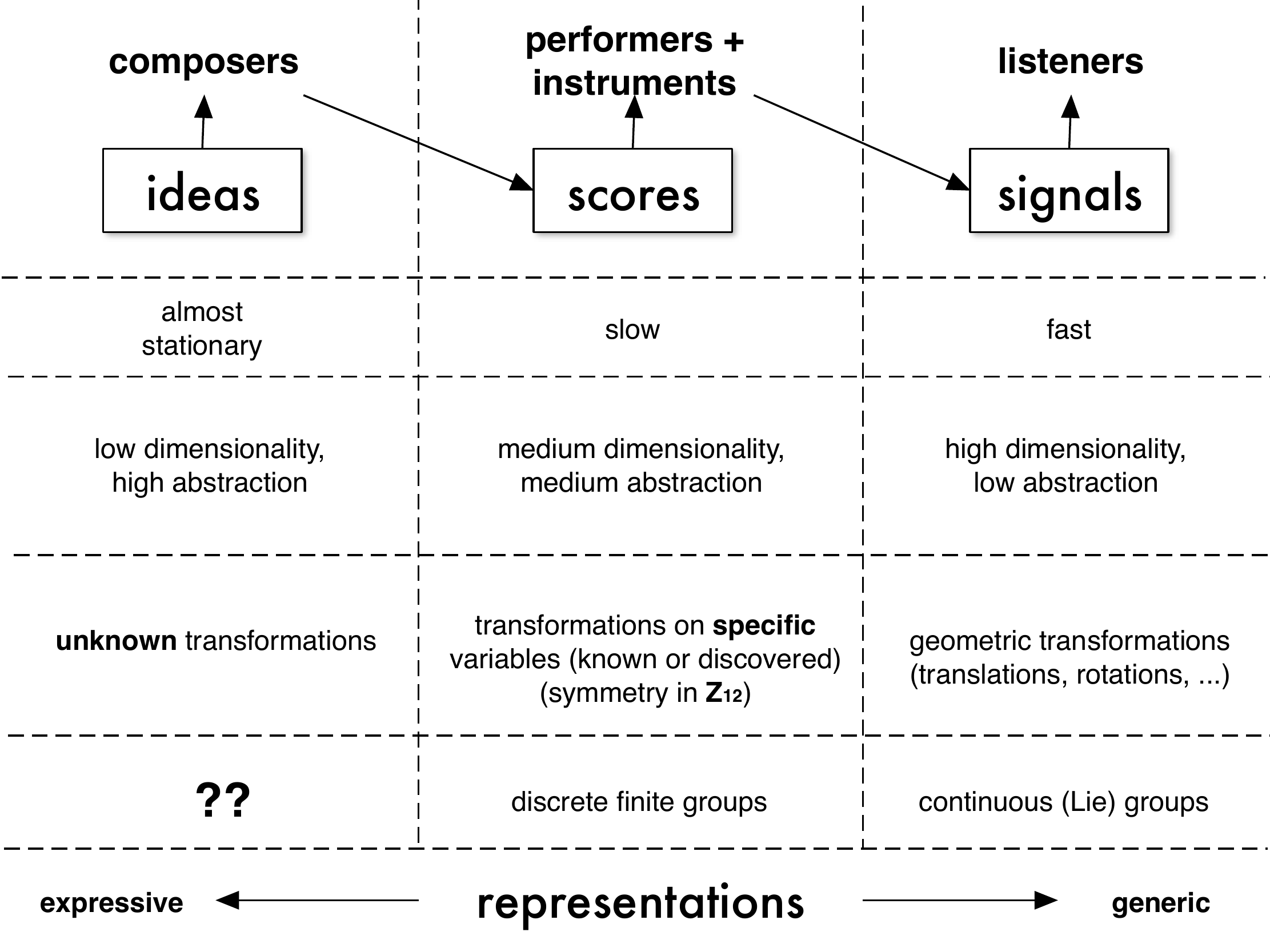}\caption{\label{fig:Different-abstraction-degrees}Different abstraction degrees
in representations.}
\end{figure}

Usually, the description of acoustic signals 
select a particular degree of abstraction in the domain of the representation.
In general, low-level representations are generic
 and have very high dimensionality. These representations evolve
fast and the only transformations that are possible
at this level are geometric (translations, rotations, etc.) and are
mostly defined on continuous (Lie) groups \citep{Mallat2016}. In the
middle, there are families of representations (often related to perceptual concepts)
that have a medium level of abstraction and 
a not so huge dimensionality and allow for transformations
on specific concept (variables), usually defined on discrete groups
\citep{Lostanlen2016}. On the other end, very abstract representations
are pretty much expressive and have a low dimensionality;
in a sense, these representations deal with almost-stationary entities
such as musical ideas and unfortunately it is very difficult to know
which mathematical structure stays behind. As an example, we could think to low-level 
representations as signals (used by listeners), to middle-level as scores 
(used by performers) and to high-level as musical ideas (used by composers).
Figure \ref{fig:Different-abstraction-degrees} depicts the outlined concepts.

Representations can be considered linear operators that need to be invariant to sources 
of unimportant variability, while being able to capture discriminative information
from signals. As such, they must respect four basic properties; being
$x$ a signal and $\Phi x$ its representation:
\begin{itemize}
\item \emph{discriminability: $\Phi x\neq\Phi y\implies x\neq y;$ }
\item \emph{stability: $\|\Phi x-\Phi y\|_{2}\leq C\|x-y\|_{2};$}
\item \emph{invariance (to group of transformations $G$): $\forall g\in G,\Phi g.x=\Phi x$;}
\item \emph{reconstruction: $y=\Phi x\iff\tilde{x}=\Phi^{-1}y.$}
\end{itemize}
Discriminability means that if the representations
of two signals are different than the two signals must be different.
Stability means that a small modification of a signal should be reflected
in a small modification in the representation and vice-versa. Invariance
to a group of transformation $G$, means that if a member of the group
is applied to a signal, than the representation must not change; reconstruction,
finally, is the possibility to go back to a signal that is \emph{equivalent
}to the original\emph{ }(in the sense of a group of transformations)
from the representation. It is possible to divide representations in two major categories:
\emph{prior }and \emph{learned}. 

\subsection{Prior and learned representations}

In prior representations, signals are decomposed using a basis that
has been defined mathematically, in order to respect (some of) the properties given above. The general model of a prior
representation is a decomposition of a signal $x$ into a linear combination
of expansion functions: $x=\sum_{k=1}^{K}\alpha_{k}g_{k} $, 
where $K$ is the dimensionality, the coefficients $\alpha_{k}$
are weights derived from an analysis stage an functions $g_{k}$ are
fixed beforehand and are used during a synthesis stage. The choice
of the decomposition functions is dependent on the particular type
of application needed and the more compact (sparse) the representation
is, the more the functions are correlated to the signal. 

In learned representations, the decomposition functions used to describe
a signal are learned by a training on some examples that belong to
a specific problem. The training can be done in two different ways:
\emph{supervised} or \emph{unsupervised}.

Supervised learning is a high-dimensional interpolation problem. We
approximate a function $f(x)$ from $q$ training samples $\lbrace x_{i},f(x_{i})\rbrace_{i\leq q}$,
where $x$ is a data vector of very high dimension $d$. A powerful
example of supervised learned representation are convolutional neural
networks.

In unsupervised training, on the other hand, there is not a target
function to approximate and other mathematical constraints are applied,
such as sparsity or variance reduction. Typical examples of unsupervised representations are sparse coding
and auto-encoders.

\subsection{Importance of unsupervised learning\label{sec:Unsupervised-learning-in}}

Recent advancements showed that supervised learning
is able, if given the right conditions, to outperform both unsupervised
and prior representations. There are problems, however, where this family of representations
cannot be applied and the only possible approach for learning
a representation is using unsupervised methods:

\begin{itemize}
\item \emph{lack of labeled data}: copyright issues can make impossible to deploy 
a database of labeled music of a significative size to be used a 
reference case for reproducible research;
\item \emph{cost of data gathering}: some real-life problems are related 
to contexts in which is impossible to gather large amount of data (such as biomedical recordings);
\item \emph{conceptual disagreement}: in some cases, it is very difficult or even 
impossible to assign labels to acoustic signals given their inherent ambiguity (music is often
such a case).

\end{itemize}

\section{Research approach}
\label{sec:researchapproach}

The discussion given above outlines, in our opinion, the necessity
of a general framework able to integrate the different approaches
to the representation of musical and acoustical signals into a common
perspective. 

\emph{Machine listening intelligence} aims at being such a framework,
by integrating cognitive musicology insights, established signal processing
techniques and more recent advancements in deep learning in the context
of \emph{representation theory}.

Prior representations are defined by mathematical models but fail
to achieve the same expressivity of learned representations. On the
contrary, deep learning proved to be valuable in incredibly different
domains and showed that some learning techniques are indeed general.
One of the issues, in the case of acoustic and musical signals, is
that there is not a common and established mathematical model for
this kind of methods. Moreover, supervised learning is not always
possible for the reasons outlined in section \ref{sec:Unsupervised-learning-in}.
Therefore, we envision a research approach based on the following
main factors:
\begin{itemize}
\item \textbf{deep unsupervised learning methods}: only with deep architectures
it is possible to create multi-scale representations that have different
abstraction levels; using unsupervised learning, it is possible
to address difficult problems that lack labelled data;
\item \textbf{representation theory}: by interpreting learning with linear
operators, it is possible to create a common mathematical language
to compare and study its properties;
\item \textbf{large scale problems}: using large datasets (that usually
embody difficult problems) will impose the research of scalable and
general learning methods, that can be transferred to many different
domains;
\item \textbf{multi-disciplinarity}: putting together several sources of
knowledge such as psychoacoustics, cognitive musicology, computational
neurobiology, signal processing and machine learning is the key for
future development of the so-called \emph{intelligent} machines.
\end{itemize}

While prior representations formally define the level of abstraction, they cannot
reach the same level of aggregate information gathered by deep learning
networks. These networks, on the other hand, are not capable of explaining
the concepts they discover. For such reasons, it is interesting to
make a bridge between these two approaches by immersing both in a
more general framework that could be found in representation theory.
Machine listening intelligence aims at filling this gap.

\bibliography{paper}

\end{document}